\documentclass[10pt]{article}
\usepackage{a4wide}
\usepackage{times}
\usepackage{subfigure}
\usepackage{epsfig}
\usepackage{float}
\usepackage{threeparttable}

\begin{document}

\title{Local Community Identification through User Access Patterns \thanks{Rodrigo Almeida and Virg\'{\i}lio Almeida and are partially supported by a number of grants from CNPq-Brazil}}

\author{Rodrigo B. Almeida~~~~~~~~Virg\'{\i}lio A. F. Almeida \\
\\
Department of Computer Science  \\
Universidade Federal de Minas Gerais \\  
Belo Horizonte, MG 31270-010 Brazil \\
\{barra,virgilio\}@dcc.ufmg.br \\
}

\date{}

\maketitle

\begin{abstract}
Community identification algorithms have been used to enhance the quality
of the services  perceived by its users. Although algorithms for community
have a widespread use in the Web, their application to portals or specific 
 subsets of the Web has not been much studied. In this paper, we propose a
 technique  for local community identification that takes
into account user access behavior derived from access logs of servers in the Web. 
The technique takes a departure from the existing  community algorithms since it changes the focus of interest, moving   from authors to users.
 Our approach does not use relations imposed by authors (e.g. hyperlinks in 
 the case of Web pages). It uses information derived from user accesses to a 
 service in order to infer relationships. The communities identified are of great
interest to content providers since they can be used to improve quality of 
their services. We also propose an evaluation methodology for
analyzing the results obtained by the algorithm. We present two case studies based on 
actual data from two services: an online bookstore and an online radio. The case of
the online radio is particularly relevant, because it emphasizes the contribution
of the proposed algorithm to find out communities in an environment (i.e., streaming
media service) without links, that represent the  relations imposed by authors 
(e.g. hyperlinks in   the case of Web pages).
\end{abstract}

\section{Introduction}
\label{sec:introduction}

Community identification algorithms have been extensively used as a way to
improve the quality perceived by users navigating through the  Web. 
Search engines have incorporated this kind of technology as a
source of information for their ranking algorithms and new applications, such
as automatic directory creation. Furthermore, community identification
studies have proven to be of great value to researchers trying to 
increase their understanding of the information 
society.~\cite{SMALLWEB,GILES1,HITS2, GILES3}. 

The use of community identification algorithms  to local communities, such
as those that interact with portals or use specific services in the  Web,
has not been much studied. The direct application of existing algorithms
to local community identification does not yield relevant results. The main reason
is the difference between the processes associated with service creation
in the two levels: local and global. The creation of services in
the Web as a whole, global context, is governed by  distributed and
uncoordinated processes. For instance, someone's decision to reference one page
authored by someone else does not have to go through any regulatory agency and
does not need its peer's authorization. Therefore, the majority
of links in the Web can be considered to have a semantic of reputation
associated with it.~\cite{PAGERANK,GILES2,HITS2, davood}. Differently, portals
are created in a centralized and coordinated manner. The structures
are created for navigational and business purposes leading to a
completely different structure~\cite{LADATHESIS}. This is why
current community identification algorithms do not provide good results
in that type of environment.

The availability of user access information in the case of a local context is
another important fact that should be noted. The combination of the community 
identification algorithms with user access information would be very valuable
to content providers, that can provide specific services to specific
communities~\cite{USERSIDE}.

 The inclusion of user access patterns on the 
community discovery process
also allows us to infer communities even from a source that does not have
explicit relationship information. Neither the books of an online bookstore nor
the games provided by an ISP are explicitly related and, therefore, can take
advantage of such technique. As the Web evolves, new kinds of services, not
explicitly related, are created and made available to the users
 accentuating the need for algorithms
designed to work based on evidences other than link information. Examples include
streaming media and game services.

This work proposes and evaluates a technique for local community
identification based on user access patterns. Our approach starts from a 
well-known community identification algorithm, the {\sl Hyperlink-Induced Topic
  Search} (HITS). We then propose a way of transforming  user access
information into a graph-based structure to be used  jointly with the HITS algorithm.  
A methodology to evaluate communities that takes into account the semantic meaning
associated with each community is also supplied. In order to exemplify the benefits of our
approach, we show  two case studies based on services available on the Internet. 

The paper is organized as follows: in Section~\ref{sec:related-work} we
present the related work. Section~\ref{sec:local-comm-ident} presents
the local community identification algorithm and proposes a methodology
to evaluate the results. Section~\ref{sec:case-studies}  presents
two case studies, based on actual logs from real online services.
Section~\ref{sec:conclusion} discusses the concluding remarks and future work.

\section{Related Work}
\label{sec:related-work}

A considerable amount of research has been developed on community
identification over the Web. Most of the approaches focus on analyzing text
content, considering vector-space models for the objects usually related to
Information Retrieval~\cite{berthier}, hyperlink structure connecting the pages~\cite{HITS2,GILES1, GILES2}, markup tags associated with the
hyperlinks or the combination of the previously cited sources of
information~\cite{enhanced,bharat}. Therefore, they are restricted to
objects that contain implicit information provided by the authors.Our work, on
the other hand, is based solely on user access behavior.

Besides, we are considering community identification applied to a local
context instead of the whole Web. Our approach aims to adapt the graph
based community identification algorithm described in~\cite{HITS2}. Some
modifications to~\cite{HITS2} that takes into account user information have
already been proposed in~\cite{mod}. However, this work was not focused on the
community identification capabilities of~\cite{HITS2} and also considered a
different representation of user patterns.

Other relevant aspect of our work is the proposal of a community evaluation
methodology that can be applied to other techniques already proposed such
as~\cite{mining,USERMACHINE} for comparison purposes. Most of the comparison
methodologies proposed so far are based on disjunction and coverage of the
communities not taking into account semantic meaning.

\section{Local Community Identification}
\label{sec:local-comm-ident}

\subsection{The HITS Algorithm}
\label{sec:hits-algorithm}

The HITS algorithm was initially proposed as a method to improve the quality of
searches on the Web~\cite{HITS1}. It takes answers to a query from a text-based
search engine and changes the ranking of these Web pages considering
the underlying hyperlinked structure connecting them. This approach,
formerly known  as link analysis, was also the base for several other
related studies~\cite{LADATHESIS,PAGERANK,GILES1}. The links are
considered as a way to represent correlations  between pages, inducing a certain
reputation/quality to a Web page pointed to by another.

The algorithm identifies pages that provide valuable information for a
determined query and also, pages that are sources of good links for
the query. These two kinds of pages are respectively called authorities and
hubs. The query in the search application is used to limit the scope of 
the Web considered by the algorithm at each execution. Therefore, 
it limits  its coverage to a certain subject expressed by a user in terms of his/her query.  

The idea behind HITS is to identify hubs and authorities 
through a mutually reinforcing relationship existent between the pages. This
relationship may be expressed as follows: a good hub is a page that
points to good authorities and a good authority is a page that is pointed to by
good hubs. This approach is very successful for the search application
since it lacks some of the weakness presented by other simple link analysis 
strategies like indegre and outdegre ranking~\cite{HITS1}.

An iterative algorithm may be used to break the circularity of the mutually
reinforcing relationship and to compute authority and hub weights for
each page. Thus, each page $p$, has associated with it an authority weight
$a_p$, and a hub weight $h_p$. These weights form a ranking of the
pages ranging from good hubs/authorities, with high $h_p$/$a_p$ values, to
bad ones, with low $h_p$/$a_p$. The weights are iteratively
evaluated by the following procedure:

\begin{eqnarray}
  a_p = \sum_{q \rightarrow p}{h_q} \nonumber \\
  h_p = \sum_{p \rightarrow q}{a_q} \nonumber
\end{eqnarray}
where $p \rightarrow q$ indicates the existence of a link link from $p$ to $q$.

Let A denote the adjacency matrix of the Web page's subgraph to be considered
by the HITS algorithm, i.e., $A[p,q]$ is equal to one if there is a link from $p$
to $q$ and 0 otherwise. The process of computing the weights may be rewritten
to:
\begin{eqnarray}
  a = {A^T} h = {A^TA} a \nonumber \\
  h = {A} a = {AA^T} h \nonumber 
\end{eqnarray}
where $a$ and $h$ are 
arrays storing authority and hub information
for all the pages considered. Then, it can be shown that, the authority and hub
arrays, $a$ and $h$, converge to the principal eigenvector of $A^T A$ and
$AA^T$ respectively.  

Although the initial work of HITS only considered the principal eigenvector of
$A^TA$ and $AA^T$, an extension to it~\cite{HITS2}, proposed to use the same
approach  to identify communities of pages over the whole Web. The approach of
the authors is to use the non-principal eigenvectors of the matrices $A^T A$ and
$AA^T$ in
order to identify other communities of Web pages. Thus, by computing the
non-principal eigenvectors we can identify other $a$'s and $h$'s arrays
identifying other communities. An implicit ranking of
the communities can be derived by this method: the principal eigenvectors
identifies the most important community over the pages, the second
principal eigenvectors explicit the second most important community over the
 pages, etc.

Our approach  applies the methodology to find communities on a graph,
introduced by Kleinberg, to another context. Our goal is to
identify communities of users that share a common interest, while accessing a
service. User access patterns are used in order to infer relationships
between them. The  generation of the graph representing the relationships and
its application to the HITS algorithm is described in the next Section.

\subsection{Community Identification Process}
\label{sec:graph-struct-gener}

Usually, the information used by the community identification
algorithms is provided by the authors of the services. Thus
the communities identified reflect the authors' perception of the
world. For instance, when these methods are applied to the
Web~\cite{GILES2,HITS2}, they use information explicited by the links
connecting the pages as a way to infer relationship between them. This kind of
information such as links or textual information is provided on
the creation of the Web page and is influenced  by the author's unique view
of the object. 

In the case of local community identification, i.e., community
identification restricted to a specific service, we consider  
the users's viewpoint. The author's point of view can be taken from 
the centralized process that creates the service, which is directly 
reflected in the service organization 
(e.g., the navigational structure of the service). 

Although user access patterns are of great interest
to local community identification, it is not straightforward how it should 
be treated. 

At first sight, we would consider the objects as being the source of a unique
view about a certain subject. This procedure is successful while considering the
whole Web since each page represents a view about a subject provided by its
author. Using objects as nodes at the graph and links derived from user access
data would not represent such a unique interpretation of the data since different
users have different interests when accessing an object. Therefore, we propose
to use the accesses to a service in order to create a graph that maps
the relationship between its users and not between its objects. This is a
departure from the traditional approach taken by  community identification
algorithms. 

Most  services in the Web  log files that record user requests to their
objects. These logs have information about the objects requested by each user
and some additional information such as the time it was issued or the status 
returned. Through the analysis of these logs, we can group requests into user sessions
 that are limited by a period of inactivity of the users~\cite{flavia, eveline}. In this
work, the sessions are considered to be the basic unit expressing  a  
user interest,  although other basic units such as the user itself or a single
request could be used with a few adaptations. 
Each session presented in the log is considered to be a node of a graph that
represent user access patterns.  The connection between any two nodes $p$ an $q$ is
directed and the weight, $S[p,q]$, between related nodes is computed by:

\begin{eqnarray}
S[p,q] = \frac{\mid O_p \cap O_q \mid}{ \mid O_p \mid} \nonumber
\end{eqnarray}
where $O_n$ represents the set of objects accessed in the nth session.

After constructing matrix $S$, that expresses the relationship
between the user sessions, we identify the communities by applying the HITS
algorithm exchanging $A$ to  $S$. The authority weight of a session $s$ in a
community $c$, given by $a_{c,s}$, is used in order to characterize the
communities. The intuition behind this procedure is that the authority weight
of a session is related to the authority weight of the objects requested in
it. The subject treated by each community is implicitly defined by its
members, i.e., sessions with high authority weights.  

\subsection{Community Evaluation and Comparison}
\label{sec:comm-eval-comp}

After user communities have been identified, a way to express their
interrelationship must be provided. This comparison, in terms of their
similarities/dissimilarities, is of great value to  service 
providers since it is this sort of information that would help them to 
design better services. For instance, one might decide to provide a personalized
service to its users based on information stored in the communities.

The weights $a_{c,s}$, associated with each pair session/community, generate a
rank of the sessions within the communities. Based on the rankings, our analysis
tries to identify good and bad sessions for each community. 
These rankings pass through a series of data analysis techniques in
order to provide interrelationships and interpretations for the communities. 

We use the Spearman rank correlation coefficient~\cite{spearman} to compare
two communities. This correlation coefficient is a non-parametric
(distribution-free) rank statistic proposed by Spearman as a measure of the
strength of the correlation between two variables through the analysis of the
rankings imposed by them.  The Spearman method can be used  to
calculate the correlation between any communities $a$ and $b$ by:
\begin{eqnarray}
    s_{a, b} & = & 1 - 6 \sum{\frac{r^2}{N(N^2 - 1)}} \nonumber
\end{eqnarray}
where $r$ is the difference in rank position of corresponding sessions. The
$s_{a,b}$ value, can be considered to be an approximation to the exact
correlation coefficient that could be found if the authority weights
for each session were considered.

The Spearman rank correlation varies from -1 to 1. Completely
opposite rankings are indicated by -1 while  equal rankings are represented 
by 1. We define distance (i.e., $d_{a,b}$) as the separation of two communities
and calculate it by the following: 
\begin{eqnarray}
  d_{a,b} = 1 - s_{a,b} \nonumber
\end{eqnarray}
The above definition is useful for visualization purposes  and for
analyzing  the communities, as shown in the examples provided in
Section~\ref{sec:case-studies}. 

Another important artifact related to community evaluation is the ability to
discover the subjects represented by each of them. A simple, yet robust,
method is to take into consideration the objects accessed by the users in each
session. We split the sessions into three disjoint sets with respect to each
community, the set of members, the set of non-members and the rest of them. The
set of members is constituted by the top {\sl n} sessions of the community
ranking. The non-members set is formed by the sessions occupying the lowest {\sl n}
positions of the ranking. The remaining sessions are included in a third set
not considered through the rest of the evaluation process. The value of $n$ should be
chosen based on the level of specificity desired or on the information
available about the objects. 

After  classifying sessions as members, non-member and indifferent, 
we proceed by evaluating positively the objects accessed by the sessions
belonging to the members' set and negatively the objects accessed by the
sessions belonging to the non-members' set. The weight associated with each
pair object/session is calculated by a measure based on a {\sl
  tf-idf}~\cite{berthier} approach, usually employed by information retrieval
techniques. 
The frequency ({\sl tf}) of an object within a session represents its
importance in the scope of the session, while the distinction capability ({\sl
  idf}) provided by the object is computed by:
\begin{eqnarray}
  idf_o = log \left( \frac{N}{N_o} \right) \nonumber
\end{eqnarray} 
where $N$ represents the total number of sessions and $N_o$ represents the
number of session in which the object $o$ was accessed.

\section{Case Studies}
\label{sec:case-studies}

This section  presents the results obtained by the application of the
proposed techniques to two different applications: an online bookstore and
an audio streaming media server providing content for an online radio. The
focus of interest are the books for the online bookstore and songs for the
audio streaming media server. The main reason for choosing these applications
was the lack of any explicit relationship between objects provided by the
service authors. 
The data comprises one week of accesses to each service. The dataset from the
bookstore was collected from August 1st to August 7th of 1999, while the
audio streaming media dataset was collected from January 13th to January 19th
of 2002. 

The online bookstore considered here is a service specialized on Computer
Science literature and operates exclusively on the Internet. Throughout the
period, the bookstore received 1.7 million requests, 50,000 of which were 
requests for information about books, such as:
authors, price, category, and reviews. Only those types of requests were 
considered in this experiment. We used 30 minutes as a threshold for the period
of inactivity. As a result, we found 40,000 users sessions.

The online radio service provides a Web interface to an audio streaming media
server that provides songs. Users can create personal radios, by specifying the
songs they want to listen to, or choose a previous stored radio. In the process
of radio creation, users can listen excerpts of the songs before they are
inserted in the radio's playlist. The streaming media server received 2.3 million
requests, 662,000 of them were requests for full songs. Only those requests were
considered in this experiment because we only wanted to capture the behavior of
users who were already listening established radios. Again we used 30
minutes of inactivity period as a threshold. The number of sessions found was
78,000. 

\subsection{Online Bookstore}
\label{sec:online-boookstore}

\begin{table*}
\scriptsize
\begin{center}
\begin{tabular}{|c|c|c|c|c|} \hline
\multicolumn{5}{|c|}{\bf{Communities}} \\ \hline
C1 & C2 & C3 & C4 & C5 \\ \hline
\multicolumn{5}{|c|}{\bf{Best-ranked categories}} \\ \hline
Certification
&Networking
&Programming
&Programming
&Databases\\ \hline
Databases
&Programming
&Networking
&Operating Systems
&Hardware\\ \hline
Reference - Education
&Web Development
&Operating Systems
&Hardware
&Digital Business \& Culture\\ \hline
\multicolumn{5}{|c|}{\bf{Worst-ranked categories}} \\ \hline
Web Development
&Reference - Education
&Microsoft
&Databases
&Reference - Education\\ \hline
Networking
&Certification central
&Databases
&Microsoft
&Programming\\ \hline
Programming
&Databases
&Certification Central
&Certification Central
&Certification Central\\ \hline

\multicolumn{5}{c}{\vspace{0.5cm}} \\ \hline

\multicolumn{5}{|c|}{\bf{Communities}} \\ \hline
 C6 & C7 & C8 & C9 & C10 \\ \hline
\multicolumn{5}{|c|}{\bf{Best-ranked categories}} \\ \hline
Certification
&Programming
&Programming
&Databases
&Certification Central\\ \hline
Microsoft
&Operating Systems
&Operating Systems
&Web Development
&Databases \\ \hline
Networking
&Web Development
&Web Development
&Operating Systems
&Reference - Education \\ \hline
\multicolumn{5}{|c|}{\bf{Worst-ranked categories}} \\ \hline
Home \& Office
&Networking
&Microsoft
&Programming
&Web Development \\ \hline
Programming
&Microsoft
&Databases
&Microsoft
&Networking \\ \hline
Databases
&Certification Central
&Certification Central
&Networking
&Programming \\ \hline

\end{tabular}
\end{center}
\caption{Qualitative analysis for the online bookstore dataset}
\label{tab:qualitative:bookstore}
\end{table*}

\begin{figure*}
\begin{center}
\mbox{
\subfigure[Sammon's mapping representation of the
communities]{\epsfig{file=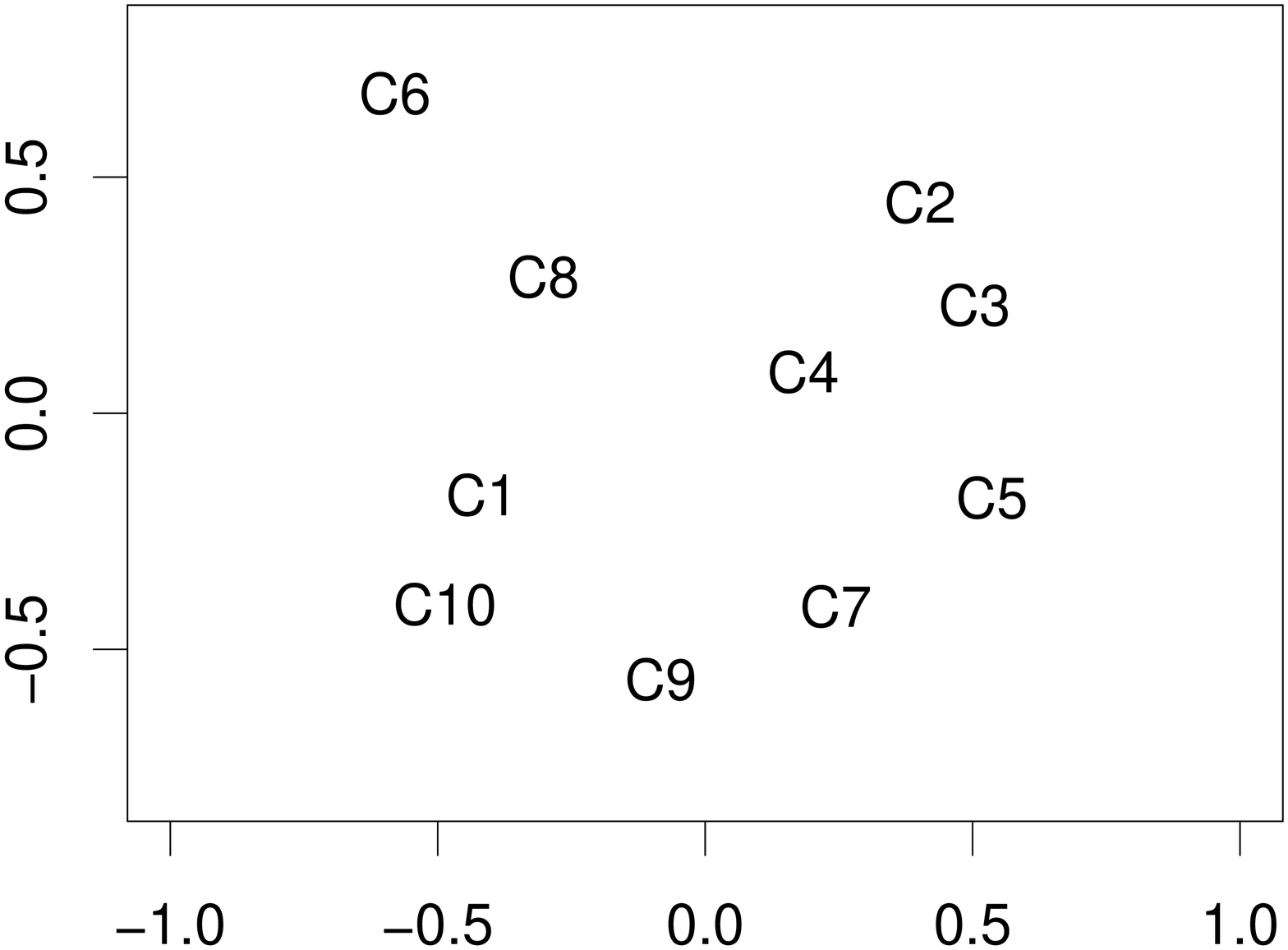, height=1.9in, angle = -90}}
\subfigure[Community cluster dendogram based on the complete method]{\epsfig{file=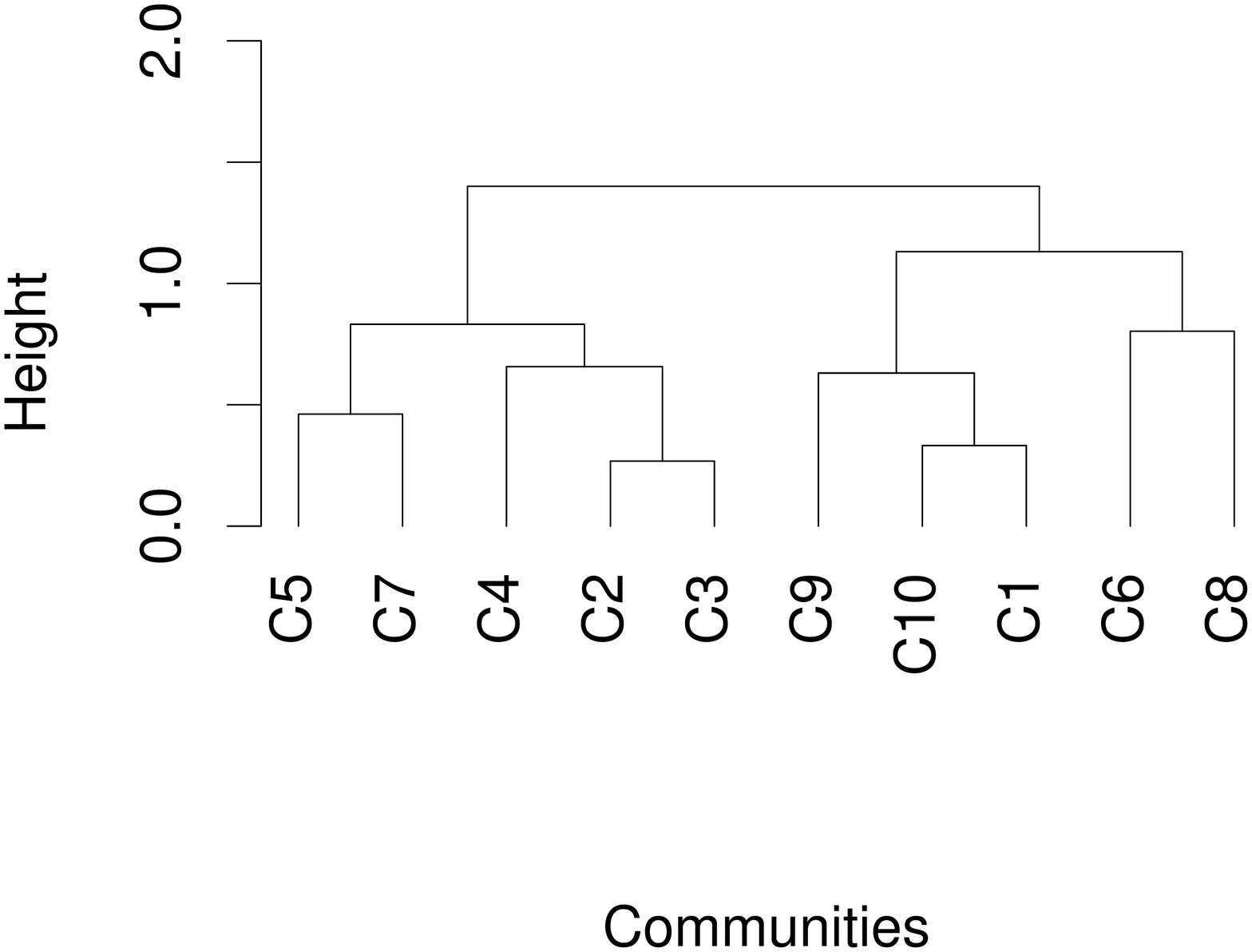, height=1.9in, angle = -90}}
\subfigure[The clusters' distance in each merge]{\epsfig{file=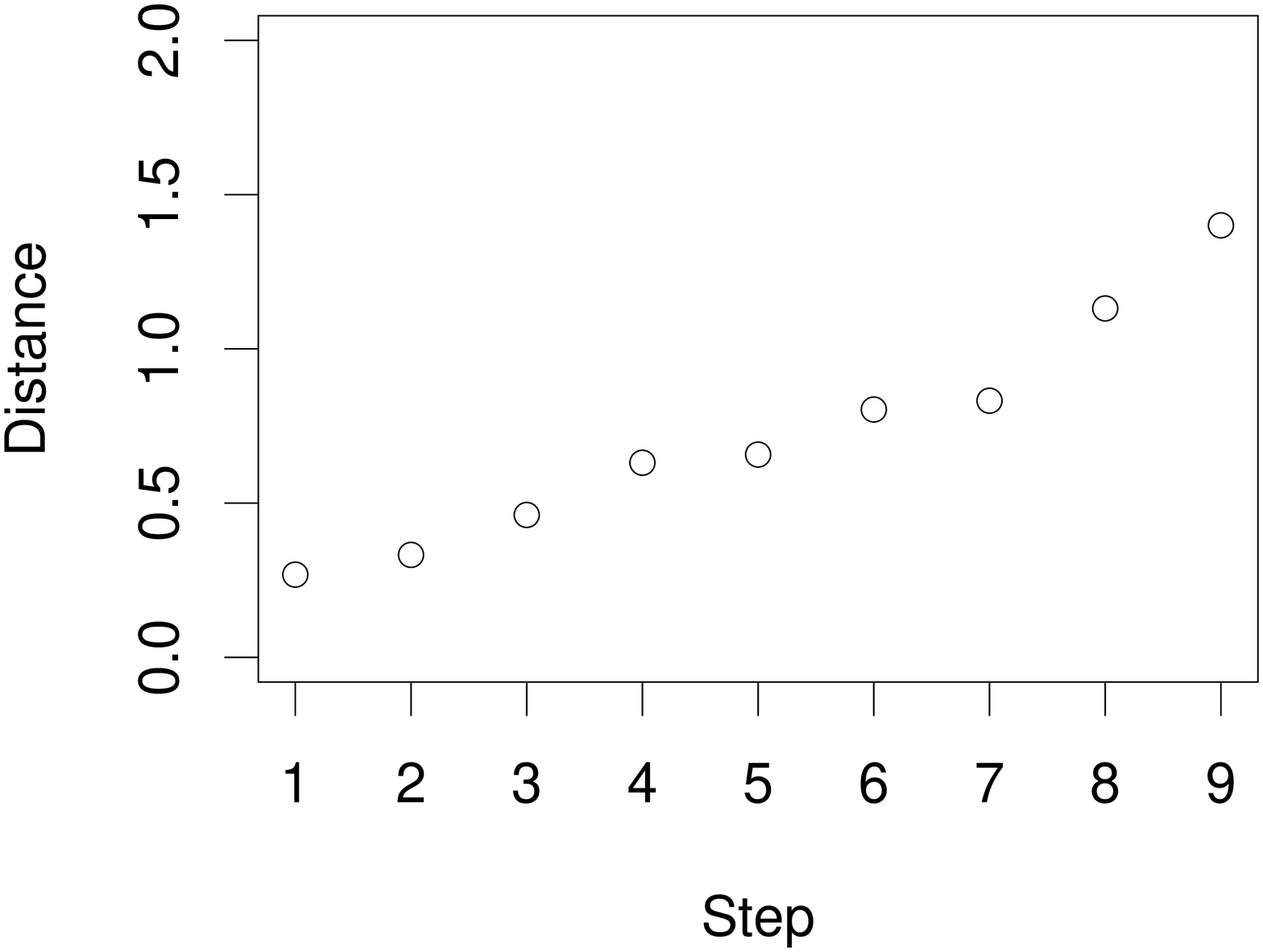, height=1.9in, angle = -90}}
}
\end{center}
\caption{Results for the online bookstore dataset}
\label{fig:bookstore:graphs}
\end{figure*}

The local community identification technique was applied to find the top 10
communities in the online bookstore dataset. The communities were named from C1 to C10.
The qualitative analysis of subjects covered by each community is presented in
Table~\ref{tab:qualitative:bookstore}. For this analysis, the first half of the
sessions' ranking were considered to be the community members and the second
half the non-members. Information about the categories that each book belongs
to were collected from the Amazon~\footnote{http://www.amazon.com} online
store. The weight of each book, computed as explained in
Section~\ref{sec:comm-eval-comp}, were used  to find the most and the least
important categories for each community.
Analyzing Table~\ref{tab:qualitative:bookstore}, we can
infer the following interpretation for  each community:
\begin{itemize}
\item Community C1 is basically formed by users interested in database
  certification that show less interest in network and programming questions
  related to Web development.
\item Community C2 have users that show some interest in Web development, while
  their main interests are networked applications involved with it, less
  importance  is given to database and certification program by the users of
  this community.
\item Distributed computing is the main interest treated by community C3, this
  community is less related to databases and Microsoft platforms/applications.
\item Community C4 aggregates users interested in low level programming
  basically related to operating system's issues. It is interesting to note that
  this community is also less related to databases and Microsoft, similar to
  community C3, because these platforms are,
  generally, less flexible. 
\item The main interests of users pertaining to community C5 are hardware
  specification of database systems. The close
  relationship between this community and management of such systems, induced by
  the interest in digital business books, also worths mention. 
\item Community C6 is formed by users interested in network administration
  of Microsoft systems. The community is mainly related to certification
  programs about this subject.
\item Low-level programming and scripting for Web development  are the main
  interests of users belonging to community C7. They are are not interested,
  although, in the network problems related to Web development. 
\item Community C8 is also related to low-level Web development issues.
  The main difference between C8 and C7 is the fact that in C8 the database
  category is underprivileged in favor of the ones bottom-ranked in C7.
\item Community C9 is also related to Web development such as C7 and C8,
  although the users of this community express some interest in certification
  programs. This information was gathered by the analysis of the whole set of
  categories for this community.
\item Certification in database systems is the main concern of users belonging
  to community C10.
\end{itemize}

We use two data analysis techniques, Sammon's mapping and Hierarchical Clustering,
to increase our understanding of the communities.  
The Sammon's mapping~\cite{sammon} is a nonlinear projection method closely
related to metric Multi-Dimensional Scaling (MDS). This method tries to
optimize a cost function that describes how well  pairwise distances in a
data set are preserved on the generated projection. The projection derived by
the use of Sammon's mapping can be seen in Figure~\ref{fig:bookstore:graphs}.(a).
Figure~\ref{fig:bookstore:graphs}.(b) shows the Hierarchical Clustering obtained
by the use of the the pairwise distance matrix between communities and the
complete-linkage method~\cite{clust}. The complete-linkage method works as
follows:  
\begin{enumerate}
\item Assign to each community its own cluster and consider the distance
  between clusters to be the same one between the respective communities.
\item Find the closest clusters and merge them into a single one. 
\item Compute the distances between this new cluster and each of the
  others. The distance between two clusters is calculated as being the longest
  distance connecting any sessions belonging to each cluster. 
\item Repeat steps 2 and 3 until all session are grouped into a single
  cluster containing all the sessions.
\end{enumerate}
Figure~\ref{fig:bookstore:graphs}.(c) shows the distances between the clusters
merged in each step of the algorithm.

As expected, both methods give similar results. They group together communities
that are 
closely related like C1 and C10, and place apart communities that
have no relation like C6 and C5. It is even more interesting to notice that 
communities like C8 and C9, that at a first look seem similar, are correctly
separated by both methods. 

The top clusters of the hierarchy shown in
Figure~\ref{fig:bookstore:graphs}.(b), separate the dataset in
 two very distinct groups. The first one, formed by (C2,
C3, C4, C5,C7) represents a group where most of the users are interested in
low-level questions, like programming and networking, usually related to
operating systems. The other one, formed by (C1, C6, C9, C10), represents a
group of users mostly interested in certification programs and their interests
vary from network administration to Web development. The dispersion of
interests found on the latter group was automatically identified. This can be
derived from the highest  cluster distance 
considered for this merge, Figure~\ref{fig:bookstore:graphs}.(c), and also by the dispersion of the
points on the Sammon's mapping projection, Figure~\ref{fig:bookstore:graphs}.(a).

The analisys of community clusters can be extended to the whole hierarchy with
similar results. The Sammon's distribution provides a comprehensive
visualization of the relations expressed by the hierarchy and the use of both
techniques together is a great start point for analysis of community data. The
quality of the results obtained in the analysis is an  evidence of the
applicability of the distance metric based on Spearman correlation. 

\subsection{Audio Streaming Media Server}
\label{sec:audio-stre-media}

\begin{figure*}
\begin{center}
\mbox{
\subfigure[Sammon's mapping representation of the
communities]{\epsfig{file=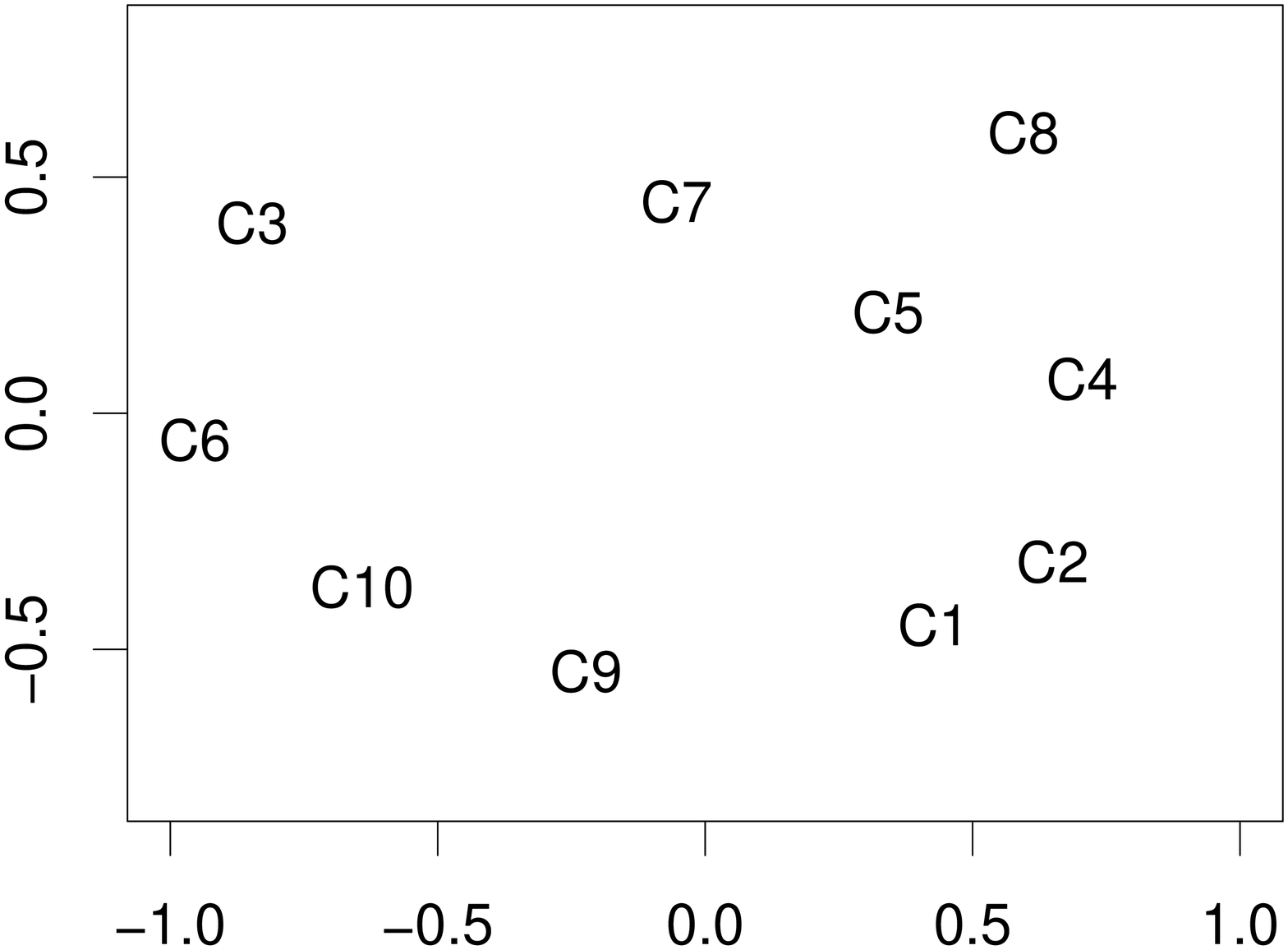, height=1.9in, angle = -90}}
\subfigure[Community cluster dendogram based on the complete method]{\epsfig{file=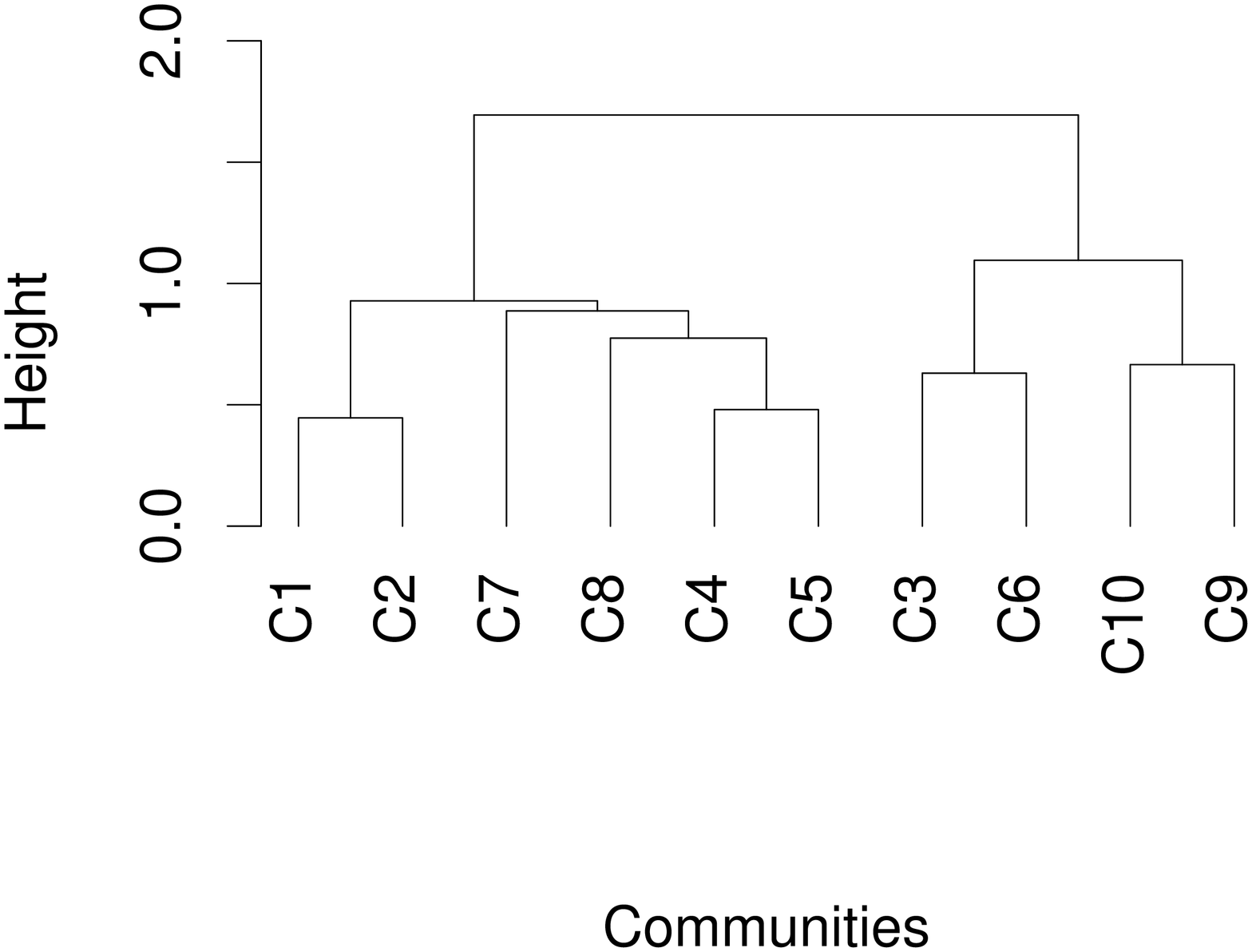, height=1.9in, angle = -90}}
\subfigure[The clusters' distance in each merge]{\epsfig{file=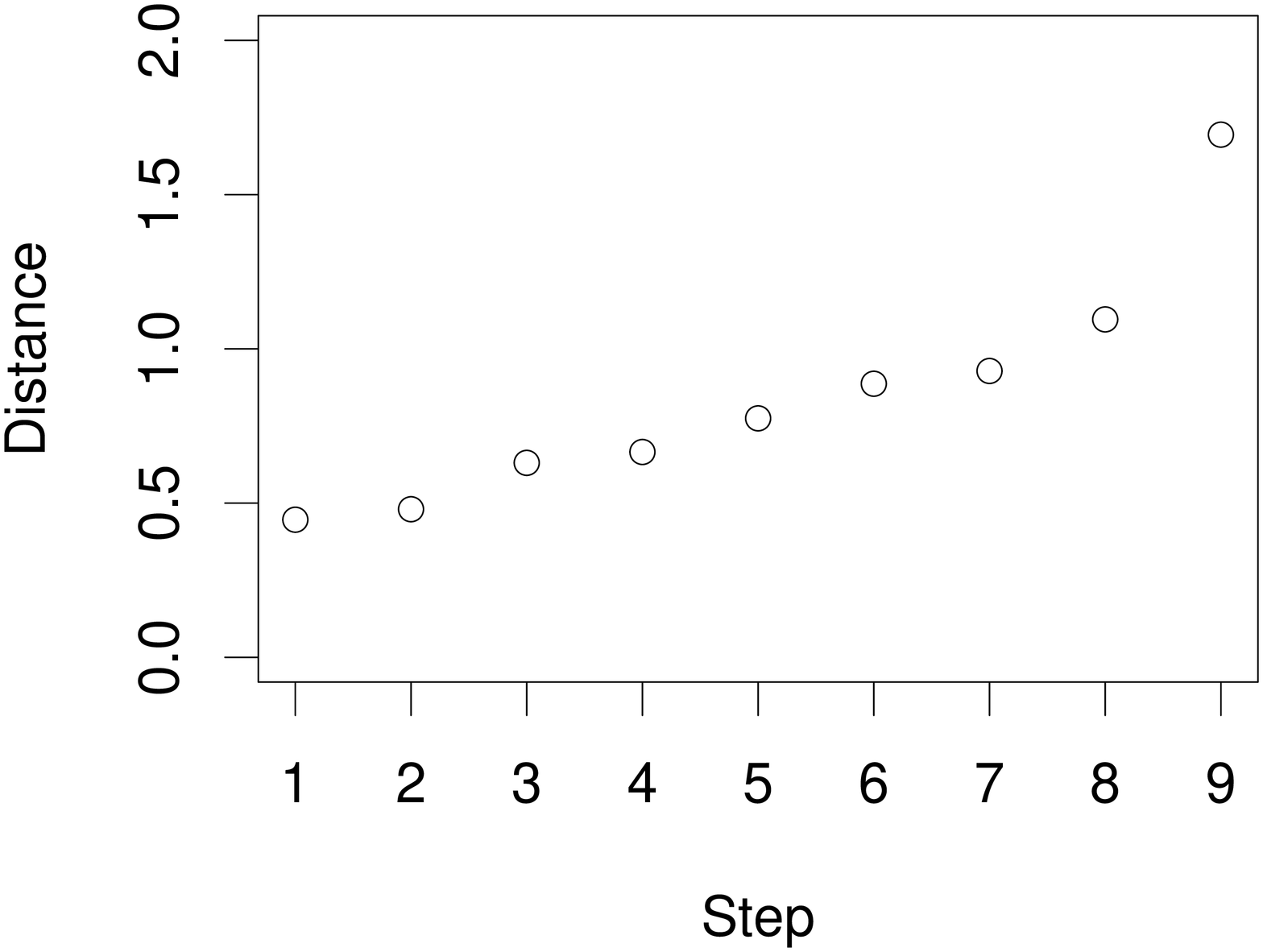, height=1.9in, angle = -90}}
}
\end{center}
\caption{Results for the audio streaming media dataset}
\label{fig:radio:graphs}
\end{figure*}

\begin{table*}
\scriptsize
  \begin{center}
    \begin{threeparttable}
      \begin{tabular}{|c|c|c|c|c|} \hline
        \multicolumn{5}{|c|}{\bf{Communities}} \\ \hline
        C1 & C2 & C3 & C4 & C5 \\ \hline
        \multicolumn{5}{|c|}{\bf{Best-ranked categories}} \\ \hline
        Brazilian Pop \& Rock
        &Samba, Ax\'{e}\tnote{1}, Pagode\tnote{2}
        &Soundtracks
        &Forr\'{o}\tnote{3}
        &Forr\'{o} \\ \hline
        Samba, Ax\'{e}, Pagode
        &Sertanejo\tnote{4}
        &International Pop
        &MPB\tnote{5}
        &Sertanejo \\ \hline
        Soundtracks
        &MPB
        &International Rock
        &International Rock
        &Brazilian Pop \& Rock\\ \hline
        
        \multicolumn{5}{|c|}{\bf{Worst-ranked categories}} \\ \hline
        MPB
        &World Music
        &Brazilian Pop \& Rock
        &Teen Pop
        &World Music \\ \hline
        World Music
        &International R\&B, Soul
        &Samba, Ax\'{e}, Pagode
        &Samba, Ax\'{e}, Pagode
        &Blues \\ \hline
        International Rock
        &International Rock
        &Sertanejo
        &Sertanejo
        &MPB\\ \hline
        
        \multicolumn{5}{c}{\vspace{0.5cm}} \\ \hline
        
        \multicolumn{5}{|c|}{\bf{Communities}} \\ \hline
        C6 & C7 & C8 & C9 & C10 \\ \hline
        \multicolumn{5}{|c|}{\bf{Best-ranked categories}} \\ \hline
        
        Forr\'{o}
        &Brazilian Pop \& Rock
        &International Pop
        &International Rock
        &International Pop \\ \hline
        Sertanejo
        &International Rock
        & Soundtracks
        &Reggae
        &International R\&B, Soul \\ \hline
        Samba, Ax\'{e}, Pagode
        &Samba, Ax\'{e}, Pagode
        &International Rock
        &Orchestras and Easy Listening
        &Soundtracks \\ \hline
        
        \multicolumn{5}{|c|}{\bf{Worst-ranked categories}} \\ \hline
        
        Brazilian Pop \& Rock
        &International R\&B, Soul
        &MPB
        &MPB
        &Sertanejo \\ \hline
        World Music
        &World Music        
        &Forr\'{o}
        &Samba, Ax\'{e}, Pagode
        &Classics \\ \hline
        MPB
        &MPB
        &Brazilian Pop \& Rock
        &Sertanejo
        &MPB \\ \hline
        
      \end{tabular}

       \begin{tablenotes}
       \item [1] Popular afro-Brazilian style from the Bahia state,
         an style closed related to carnival
       \item [2] Popular Brazilian style derived from Samba
       \item [3] Catchy dance music from the Northeast of Brazil
       \item [4] Brazilian country music
       \item [5] Commonly used for Brazilian pop coming after the Bossa
         Nova style
       \end{tablenotes}

    \end{threeparttable}
    
  \end{center}
   
  
  \caption{Qualitative analisys for the audio streaming media dataset}
  \label{tab:qualitative:radio}
\end{table*}

The same methodology used in the previous section was 
applied to the audio streaming media dataset.
The top 10 communities  (C1 to C10) existent on the dataset were
identified. The qualitative analysis of styles covered by each community and a
short explanation of some Brazilian music styles are presented in 
Table~\ref{tab:qualitative:radio}. For this analysis, only the
top and bottom 100 session were considered to be the elements of the members'
and non-members' sets. Unlike the online bookstore, we did not
have access to a unique identifier for the songs played. The  information
available was the title of the songs, CDs and artists, making the process
of categorizing  data a time-consuming task. Data about
the styles of the songs accessed on the considered sessions were
collected from Amazon and Submarino\footnote{http://www.submarino.com.br}, a
major online store in Brazil. The Sammon's mapping for this dataset is
presented in
Figure~\ref{fig:radio:graphs}.(a). Figure~\ref{fig:radio:graphs}.(b) and
Figure~\ref{fig:radio:graphs}.(c) the results obtained by the Hierarchical
Clustering method when applied to this dataset.

Like in Section~\ref{sec:online-boookstore}, we have the following
obserrvations for the identified communities. For example,  users in communities
 C9 and C10  represent users interested in international music styles,
that do   not pay much attention to Brazilian music. Communities C3 and C10,
 that are located apart in both representations, seems to represent different 
interests of their users. 
Although the same kind of analisys based on similarities of communities and
their interests can be done for this dataset, we want to point out other
dataset features identified by the algorithm without relying upon any
explicit information. One of them is related to the structure of the phonographic
industry existing in Brazil and the other one is related to the specificity of
each dataset.

Even for an untrained observer,  Table~\ref{tab:qualitative:radio} 
shows that users of the online radio exhibit strong interest in local music. Much
of the categories cited are of Brazilian music, even though all top
international albums were also available. This fact is extremely
important since it reflects what happens everyday on Brazilian
streets. The IFPI Music Piracy Report~\cite{piracy} shows that over 50\% of the
piracy in Brazil is domestic and, therefore, many questions concerning the
survival of the local phonographic industry production are being raised.
The algorithm's capability of confirming a behavior observed in the the society 
is very interesting since it can shed light on new questions.

The specificity level of each dataset is  different and the algorithm
is able to reflect this fact. The online bookstore is specialized in Computer Science
while the online radio service provides access to different music styles from
different nationalities. The slightly higher distance measures used in the
each merge step is an evidence of the latter,
Figures~\ref{fig:bookstore:graphs}.(c) and ~\ref{fig:radio:graphs}.(c). Also we
can see in the Sammon's mapping  that the communities found in
the audio streaming media dataset, Figure~\ref{fig:radio:graphs}.(a), are more
spread than the ones of the bookstore dataset, Figure~\ref{fig:radio:graphs}.(a).

\section{Concluding Remarks}
\label{sec:conclusion}

The methodology proposed offers several advantages over the graph-based
algorithms in their pure form when applied to the context of local community
identification. The communities identified represent the user's perception of
the information provided by the services, and  this understanding gives
service providers a great opportunity to service improvement. 

An evaluation methodology based on data analysis available was also
proposed. The evaluation technique is based on  {\sl tf-idf} ranking of 
occurrences and  the Spearman rank correlation. The former is used to provide
the focus of each community and the latter, derive a pairwise distance
metric. The benefits of these methods are exemplified by the case studies, based
on actual data of two real services available in the Web.

The results obtained in this paper are encouraging
and show that the proposed techniques and metrics are promising 
for characterizing the interests of users accessing  a service in the Web.
 Yet, this is
just an introductory study and we must devote much attention to other possibles
metrics, datasets and applicabilities of the proposed technique. The temporal
emergence of communities and their evolution is also of great interest. We also
intend to compare our results with other methods used for similar purposes.

\section{Acknowledgments}
The authors would like to thank the anonymous service owners and operators for
enabling this research to proceed by providing us access to their logs. We 
also would like to thank the e-SPEED lab team, at the DCC/UFMG, for the
continuous help during the development of this research, specially Fl\'avia
Ribeiro for helping with the revision of some of the text presented in this paper.



\bibliographystyle{abbrv}

\begin{thebibliography}{10}

\bibitem{SMALLWEB}
L.~A. Adamic.
\newblock The small world web.
\newblock In {\em Proceedings of the Third European Conference on Digital
  Libraries}, 1999.

\bibitem{LADATHESIS}
L.~A. Adamic.
\newblock {\em Network Dynamics: The World Wide Web}.
\newblock PhD thesis, Stanford University, 2001.

\bibitem{berthier}
R.~Baeza-Yates and B.~Ribeiro-Neto.
\newblock {\em Modern Information Retrieval}.
\newblock Addison-Wesley, 1999.

\bibitem{bharat}
K.~Bharat and M.~R. Henzinger.
\newblock Improved algorithms for topic distillation in a hyperlinked
  environment.
\newblock In {\em Proceedings of the 21st international conference on Research
  and development in information retrieval (SIGIR)}, 1998.

\bibitem{PAGERANK}
S.~Brin and L.~Page.
\newblock The anatomy of a large-scale hypertextual web search engine.
\newblock In {\em Proceeding of the Ninth International World wide Web
  Conference}, 1998.

\bibitem{mining}
S.~Chakrabarti and Y.~Batterywala.
\newblock Mining themes from bookmarks.
\newblock In {\em ACM SIGKDD Workshop on Text Mining}, 2000.

\bibitem{enhanced}
S.~Chakrabarti, M.~M. Joshi, and V.~B. Tawde.
\newblock Enhanced topic distillation using text, markup tags, and hyperlinks.
\newblock In {\em Proceedings of the 20th international conference on Research
  and development in information retrieval (SIGIR)}, 2001.

\bibitem{GILES1}
G.~W. Flake, S.~Lawrence, and C.~L. Giles.
\newblock Efficiente identification of web communities.
\newblock In {\em Proceedings of the Sixth International Conference on
  Knowledge Discovery and Data Mining}, 2000.

\bibitem{GILES2}
G.~W. Flake, S.~Lawrence, C.~L. Giles, and F.~M. Coetzee.
\newblock Self-organization of the web and identification of communities.
\newblock {\em I{E}{E}{E} Computer}, 2002.

\bibitem{HITS2}
D.~Gibson, J.~M. Kleinberg, and P.~Raghavan.
\newblock Inferring web communities from link topology.
\newblock In {\em Proceeding of the ninth Conference on hypertext and
  Hypermedia}, 1998.

\bibitem{piracy}
{IFPI} {M}usic {P}iracy {R}eport.
\newblock {\sl http://www.ifpi.org/}.

\bibitem{HITS1}
J.~M. Kleinberg.
\newblock Authoritative sources in a hyperlinked environment.
\newblock {\em Journal of the ACM}, 1997.

\bibitem{spearman}
E.~L. Lehmann and H.~J.~M. D'Abrera.
\newblock {\em Nonparametrics: Statistical Methods Based on Ranks}.
\newblock Prentice Hall College Div, 1998.

\bibitem{flavia}
D.~A. Menasc\'{e}, V.~A.~F. Almeida, R.~H. Riedi, F.~P. Ribeiro, R.~L.~C.
  Fonseca, and W.~{Meira Jr.}
\newblock In search of invariants for e-business workloads.
\newblock In {\em Proceeding of the Second {ACM} Conference on Electronic
  Commerce}, 2000.

\bibitem{mod}
J.~C. Miller, G.~Rae, F.~Schaefer, L.~A. Ward, T.~LoFaro, and A.~Farahat.
\newblock Modifications of kleinberg's hits algorithm using matrix
  exponentiation and web log records.
\newblock In {\em Proceedings of the 24th international conference on Research
  and development in information retrieval (SIGIR)}, 2001.

\bibitem{clust}
F.~Murtagh.
\newblock Multidimensional clustering algorithms.
\newblock {\em COMPSTAT Lectures 4}, 1985.

\bibitem{USERMACHINE}
G.~Paliouras, C.~Papatheodorou, V.~Karkaletsis, and C.~Spyropoulos.
\newblock Discovering user communities on the internet using unsupervised
  machine learning techniques.
\newblock {\em To appear in Interacting with Computers}, 2002.

\bibitem{USERSIDE}
G.~Paliouras, C.~Papatheodorou, V.~Karkaletsis, C.~Spyropoulos, and
  V.~Malaveta.
\newblock Learning user communities for improving the services of information
  providers.
\newblock In {\em European Conference on Research and Advanced Technology for
  Digital Libraries}, 1998.

\bibitem{GILES3}
D.~Pennock, G.~W. Flake, S.~Lawrence, E.~Glover, and C.~L. Giles.
\newblock Winners don't take all: Characterizing the competition for links on
  the web.
\newblock {\em Proceedings of the National Academy of Sciences}, 2002.

\bibitem{davood}
D.~Rafiei.
\newblock What is this page know for? {C}omputing web page reputations.
\newblock In {\em Proceeding of the Ninth International World wide Web
  Conference}, 2000.

\bibitem{sammon}
J.~W. Sammon~Jr.
\newblock A nonlinear mapping for data structure analysis.
\newblock {\em IEEE Transactions on Computers}, C-18(5), 1969.

\bibitem{eveline}
E.~A. Veloso, V.~A.~F. Almeida, W.~{Meira Jr.}, A.~Bestavros, and S.~Jin.
\newblock A hierarchical characterization of a live streaming media workload.
\newblock In {\em Proceedings of the ACM Internet Measurment Workshop
  (IMW'02)}, 2002.

\end{thebibliography}

\end{document}